\documentclass[a4paper,11pt]{article}
\pdfoutput=1 

\usepackage{jinstpub} 

\title{\boldmath The CBM Time-of-Flight system}

\author[a,1]{I. Deppner,\note{Corresponding author.}}
\author[a]{N. Herrmann,}

\affiliation[a]{Physikalisches Institut der Universit\"at Heidelberg, Im Neuenheimer Feld 226, Heidelberg, Germany}

\emailAdd{deppner@physi.uni-heidelberg.de}

\abstract{The Compressed Baryonic Matter spectrometer (CBM) is a future fixed-target heavy-ion experiment located at the Facility for Anti-proton and Ion Research (FAIR) in Darmstadt, Germany. The key element in CBM providing hadron identification at incident beam energies between 2 and 11 AGeV (for Au-nuclei) will be a 120 m$^2$ large Time-of-Flight (ToF) wall composed of Multi-gap Resistive Plate Chambers (MRPC) with a system time resolution better than 80 ps. Aiming for an interaction rate of 10 MHz for Au+Au collisions the MRPCs have to cope with an incident particle flux between 0.1~kHz/cm$^2$ and 100~kHz/cm$^2$ depending on their location. Characterized by granularity and rate capability the actual conceptual design of the ToF-wall foresees 6 different counter granularities and 4 different counter designs. In order to elaborate the final MRPC design of these counters several heavy-ion in-beam and cosmic tests were performed. In this contribution we present the conceptual design of the TOF wall and in particular discuss performance results of full-size MRPC prototypes.}


\collaboration[c]{on behalf of the CBM collaboration}

\proceeding{XIV Workshop on Resistive Plate Chambers and related detectors$^{\text{th}}$ \\
  19 - 23 February, 2018,\\
   Puerto Vallarta, Jalisco state, Mexico}

\usepackage{hyperref}

\begin{document}
\maketitle
\flushbottom

\section{Introduction}
\label{sec:intro}

The Compressed Baryonic Matter spectrometer (CBM) is a future fixed-target heavy-ion experiment located at the Facility for Anti-proton and Ion Research (FAIR) in Darmstadt, Germany. The CBM collaboration aims to explore the phase diagram of strongly interacting matter at high baryon densities in the beam energies range from 2 A GeV to 11 A GeV. This includes the study of the equation-of-state of nuclear matter at neutron star core densities, and the search for phase transitions, chiral symmetry restoration, and exotic forms of (strange) QCD matter \cite{Ablyazimov:2017guv}. The CBM detector is designed to measure the collective behavior of hadrons, together with rare diagnostic probes such as multi-strange hyperons, charmed particles and vector mesons decaying into lepton pairs with unprecedented precision and statistics. Most of these particles will be studied for the first time in the FAIR energy range. In order to achieve the required precision, the measurements will be performed at reaction rates up to 10 MHz. This requires very fast and radiation hard detectors, a novel data read-out and analysis concept including free streaming front-end electronics, and a high performance computing cluster for online event selection \cite{CBM_webpage}.

\section{The CBM Time-of-Flight (TOF) system}
\label{sec:TOF}

\begin{figure}[htbp]
	\centering 
	\includegraphics[width=1.0\textwidth]{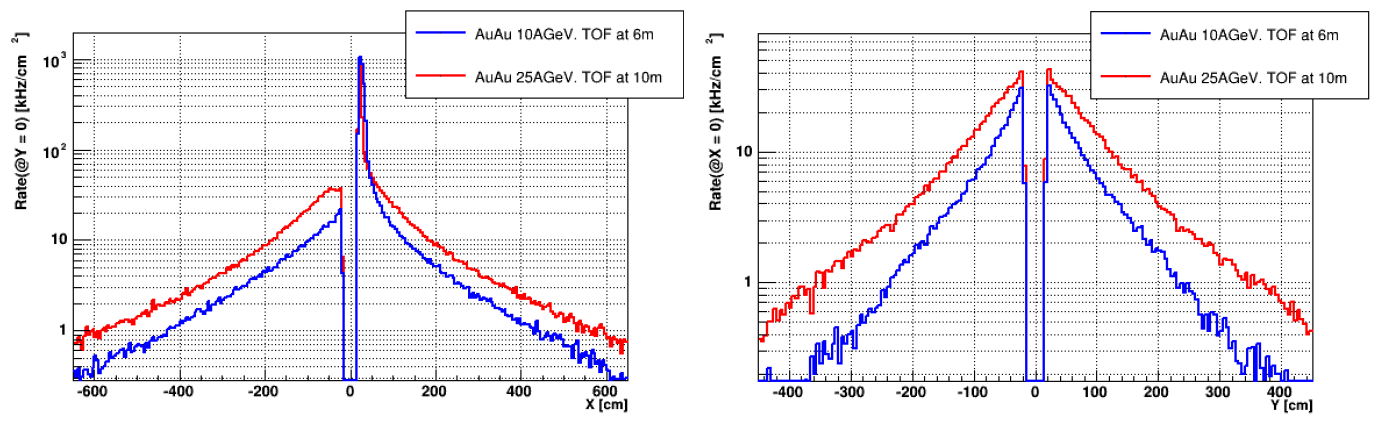}
	\caption{\label{fig:1} Calculated particle flux in the CBM TOF wall placed 10 m (red) and
		6 m (blue) behind the primary interaction point using as event generator URQMD
		and a target interaction rate of 10 MHz minimum bias Au+Au reactions at two
		different incident energies, depicting the running conditions at SIS300 (red) and
		SIS100 (blue). The particle flux includes the contribution of secondary particles
		produced in the upstream material of CBM. Left pannel: Flux as function of X at
		Y=0 (i.e. left/right of the beam axis), the X-axis also defines the deflection plane of
		the particles trajectories in the magnetic field. Right pannel: Flux as function of Y
		at X=0.}
\end{figure}

The CBM Time-of-Flight (TOF) system \cite{Herrmann:2014} will provide charged hadron identification (protons, kaons and pions) up to a particle momentum of about 4 GeV/c. In order to fulfill this requirements reliably well system time resolution of below 80 ps and an overall efficiency above 95\% is mandatory. However, the most challenging requirement for the detector is a rate capability of up to 30 kHz/cm$^2$ (confirm figure \ref{fig:1}) which occurs at the aperture of 2.5 $^\circ$. Below 2.5 $^\circ$ no information on the momentum is available. However, it is planed to install a Beam Fragmentation T0 Counter (BFTC) delivering the start time information which has to cope with particle fluxes up to 100 kHz/cm$^2$. The current conceptual design foresees a 120 m$^2$ big TOF wall (including BFTC) composed of Multi-gap Resistive Plate Chambers (MRPC) located between 6 m and 10 m from the interaction point. Since an occupancy below 5\% is desired the granularity of the MRPCs varies between 4 cm$^2$ and 50 cm$^2$ depending on their location in the TOF wall.  

The BFTC is composed of 8 modules containing 400 single ceramic MRPCs each arranged around the beam axes. The active area of a MRPC is 2 $\times$ 2 cm$^2$ while it is composed of 3 stacked cells with 2 gaps each. The gap size is 250 $\mu$m. More information about the counter design and test beam results can be found in \cite{Naumann:2011, Akindinov:2017, Sultanov:2014, Naumann:2018}. Above an aperture of 2.5 $^\circ$ glass MRPCs are foreseen. The high and intermediate rate regions (between 30 kHz/cm$^2$ and 1 kHz/cm$^2$) will be composed of MRPCs equipped with low resistive glass while below 1 kHz/cm$^2$ float glass MRPCs will be used. Details on the conceptual design of the wall can be found in \cite{Deppner:2012zz, Deppner:2014sua}. Various full size MRPC prototypes including test results foreseen for the high, intermediate and low rate region of the wall will be discussed in the following subsections.

\subsection{The MRPC1 prototype for the high rate region}

\begin{figure}[htbp]
	\centering 
	\includegraphics[width=0.8\textwidth]{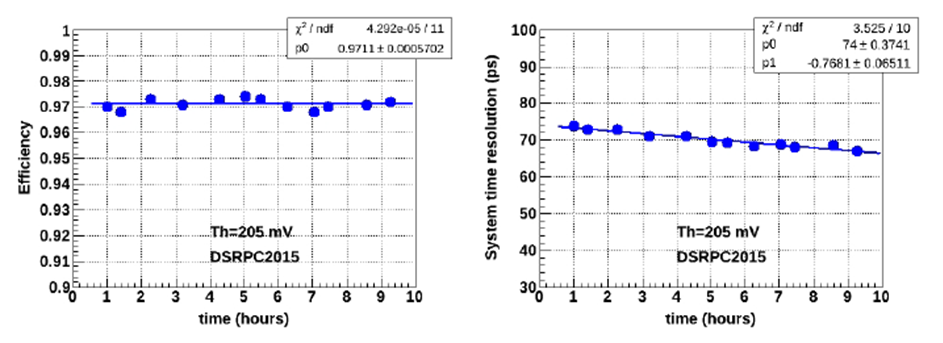}
	\caption{\label{fig:3} Efficiency (left) and system time resolution (right) obtained in a long term run of 10 hours. }
\end{figure}

\begin{figure}[htbp]
	\centering 
	\includegraphics[width=0.8\textwidth]{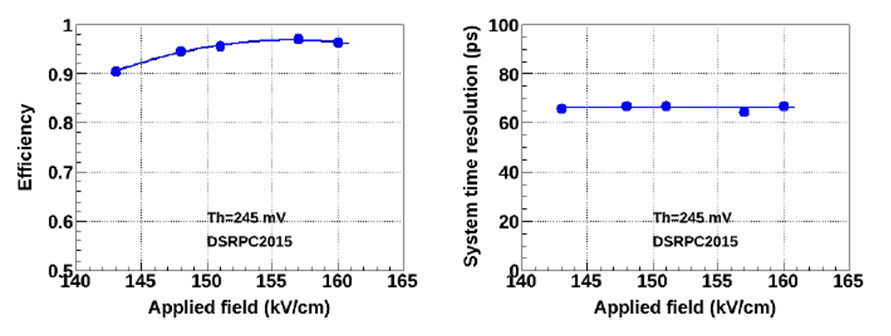}
	\caption{\label{fig:4} Efficiency (left) and system time resolution (right) as funtion of the applied el. field.}
\end{figure}

The high rate region is defined as the part where fluxes between 30~kHz/cm$^2$ and 5~kHz/cm$^2$ will be reached. Here MRPCs with granularities between 6~cm$^2$ and 20~cm$^2$ are foreseen. The MRPC1 prototype developed in Bucharest (described in \cite{Petris:2017, Petris:2018}) is equipped with low resistive glass of 0.7~mm thickness and has a granularity of about 6~cm$^2$. 
\begin{figure}[htbp]
	\centering 
	\includegraphics[width=1.0\textwidth]{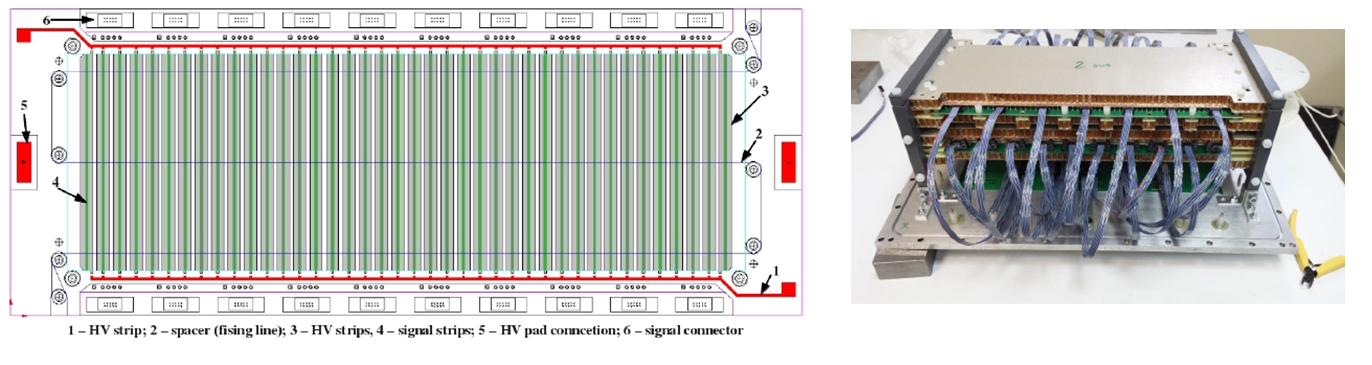}
	\caption{\label{fig:2} Left side: structure of the MRPC1 prototype developed at Bucharest. The HV electrode is consists of metal stripes with the same pitch as the readout electrode. Right side: Photograph of 2 MRPC1 prototypes called DSMRPC2015 and superimposed a different prototype called SSMRPC2015.}
\end{figure}
Figure \ref{fig:2} depicts the structure and a photograph of the MRPC1 prototype called DSMRPC2015. A novel method was developed to adapt the impedance of a MRPC independently from its read out structure \cite{Bartos:2018}. That this method works was demonstrated by the first time on this detector where the impedance was adjusted to 100~$\Omega$. The impedance matching is achieved by manufacturing a high voltage (HV) electrode from metal (e.g. a PCB) which is subdivided in strips with the same strip pitch like the overlaying read out electrode. However, the strip width of both electrodes (HV and read out) can differ. The granularity of the detector is now defined by the HV electrode structure while the impedance is fixed by the ration between the two strip widths. This prototype was tested among others at SPS/CERN with 30 AGeV Pb ions impinging on a 4 mm thick Pb target. The spray of reaction products takes care for full illumination of the detector area with a gradient towards the beam axis. The gas mixture was 85~\% R134a, 10~\% SF$_6$ and 5~\% i-Butane. Figure \ref{fig:3} shows the efficiency (left side) and the time resolution (right side) in a long term run of about 10 hours in the beginning of the beam time at modest particle fluxes of about few 100~Hz/cm$^2$. While the efficiency of about 97~\% stays constant a significant decrease in the system time resolution from 72~ps to 69~ps is observed. As a reference a second MRPC (also seen in Fig. \ref{fig:2} right side) was used. The observed behavior is considered to be a conditioning effect. After a day of operation the system time resolution becomes stable. 
The efficiency and the system time resolution as function of the applied el. field is depicted in Fig.~\ref{fig:4}. The efficiency enters a plateau at about 153~kV/cm and reaches a value of 97~\%. The high el. field is necessary since the gap size of this MRPC is only 140~$\mu$m. The system time resolution is constant and has a value of 67~ps. Assuming a equal performance of both MRPCs this value can be divided by $\sqrt{2}$ in order to obtain the individual counter resolution which is about 47~ps including the jitter of the electronics chain. 
These are very encouraging results since hit multiplicities of an average of about 5 was measured \cite{Petris:2018zvt} and confirmed in MC simulation. 

\subsection{The MRPC3a prototype for the intermediate rate region}

The intermediate rate region is defined as the part where fluxes between 5~kHz/cm$^2$ and 1~kHz/cm$^2$ will be reached. Here MRPCs with strip geometries of 1 $\times$ 27~cm$^2$ are foreseen. The MRPC3a prototype developed at Tsinghua university has 32 readout strips and a low resistive glass stack of 2 $\times$ 4 gaps with 250~$\mu$m gap size. The glass has a bulk resistivity of about 2 $\times$ 10$^{10}$ $\Omega/\Box$ and a thickness of 0.7~mm. MRPC3a is impedance matched to 50~$\Omega$. Beam time results a SIS18 for prototypes with similar structure were presented in \cite{Deppner:2016yku, Wang:2016bsx}. For the FAIR phase 0 program (see chapter \ref{sec:FAIR_phase0}) 72 MRPC3a prototypes are being produced providing a solid basis on which the quality assurance procedure can be exercised.

Extensive cosmic rays test of the MRPC3a prototype together with the MRPC3b prototype were performed in Heidelberg. The results will be discussed in the next subchapter.

\subsection{The MRPC3b prototype for the low rate region}

The low rate region is the area of the TOF wall which has to cope with particle fluxes below 1~kHz/cm$^2$. This area will be covered by MRPCs equipped with commercial thin float glass (0.28~mm) as electrode material. Foreseen are two different versions which differ only in the strip length of the pickup electrode (MRPC3b has a strip length of 27 ~cm and MRPC4 has a strip length of 50 ~cm). Both prototypes are developed and built at University of Science and Technology of China (USTC). The detectors have 2 $\times$ 5 gaps with a gap size of 230~$\mu$m and 32 readout channels with strips of 1~cm pitch. Also this prototype is impedance matched to 50~$\Omega$ which is adjusted on the preamplifier stage with a resistor. For the FAIR phase 0 program (see chapter \ref{sec:FAIR_phase0}) 80 MRPC3b prototypes are being produced.

As stated above these prototypes were tested with cosmics in a stack of 6 stations (2 MRPC3a and 4 MRPC3b detectors) in Heidelberg. Per day about 100000 tracks which traverse all 6 stations could be accumulated. The gas mixture was 90~\% R134a, 5~\% SF$_6$ and 5~\% i-Butane. The system is self triggered and the data flow is free streaming as it is foreseen in the CBM experiment. The FEE consists of the PADIX preamplifier/discriminator chip \cite{Ciobanu2014} and the self-triggered GET4 TDC \cite{Deppe:2009rpa}. The data unpacker builds events when at least 4 signals from different detectors were found. The data are calibrated regarding time and position offsets and time walk. After calibration a track finding algorithm is applied and the resulting tracks are used to determine efficiency, time and position resolution. The efficiency of each counter is determined by comparing tracks with multiplicity 5 to those with multiplicity 6. With other words, 5 counters define a track (acting as a reference) which is inter-/extrapolated to the detector under test (DUT). When a hit on DUT is found close in space and time the counter is considered to be efficient. 
Further more, residuals between the hit measured on the DUT and the track formed by the other 5 hits, can be determined. The width of the Gaussian shaped residuals deliver the time and spacial resolution of each counter. This method has the advantage that a multi differential analysis can be performed and by construction the incident angle and the velocity spread of the cosmic rays are taken into account.        
 
\begin{figure}[htbp]
	\centering 
	\includegraphics[width=.4\textwidth]{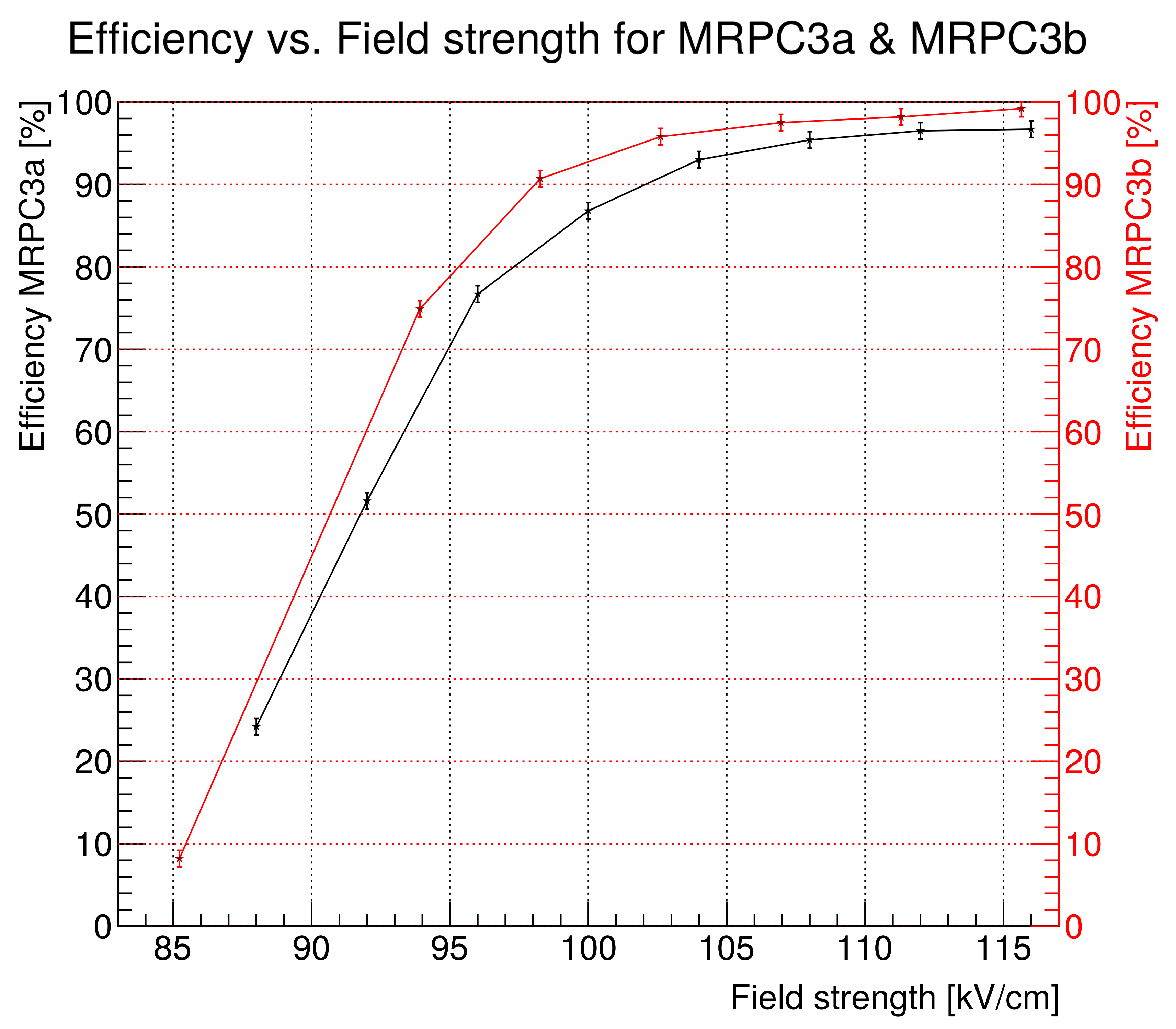}
	\qquad
	\includegraphics[width=.4\textwidth]{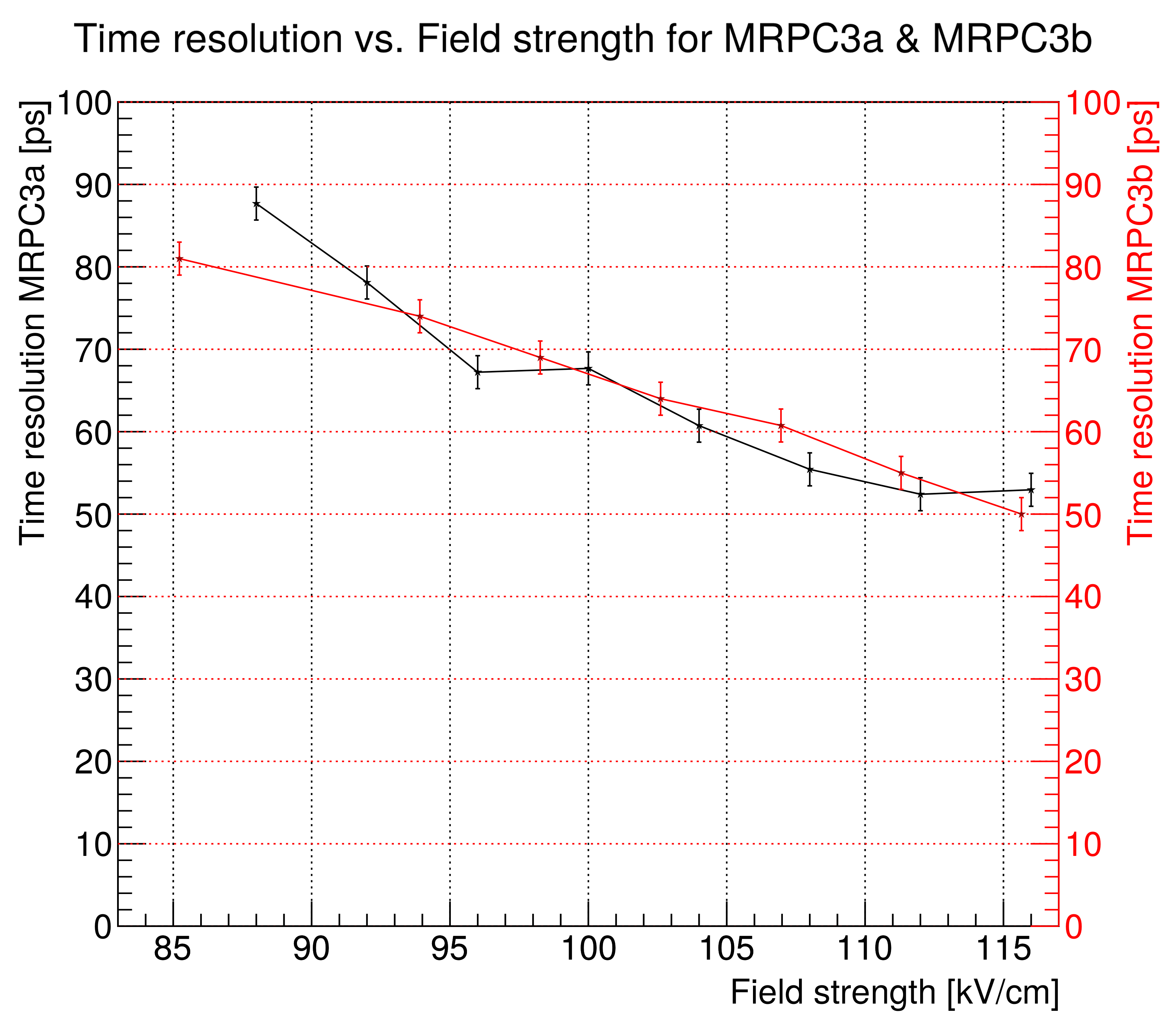}
	\caption{\label{fig:5} Left side: efficiency as a funtion of the el. field for the prototypes MRPC3a (low resistive glass) and MRPC3b (thin float glass). Right side: counter time resolution as a funtion of the el. field for the prototypes MRPC3a and MRPC3b}
\end{figure}

\begin{figure}[htbp]
	\centering 
	\includegraphics[width=.4\textwidth]{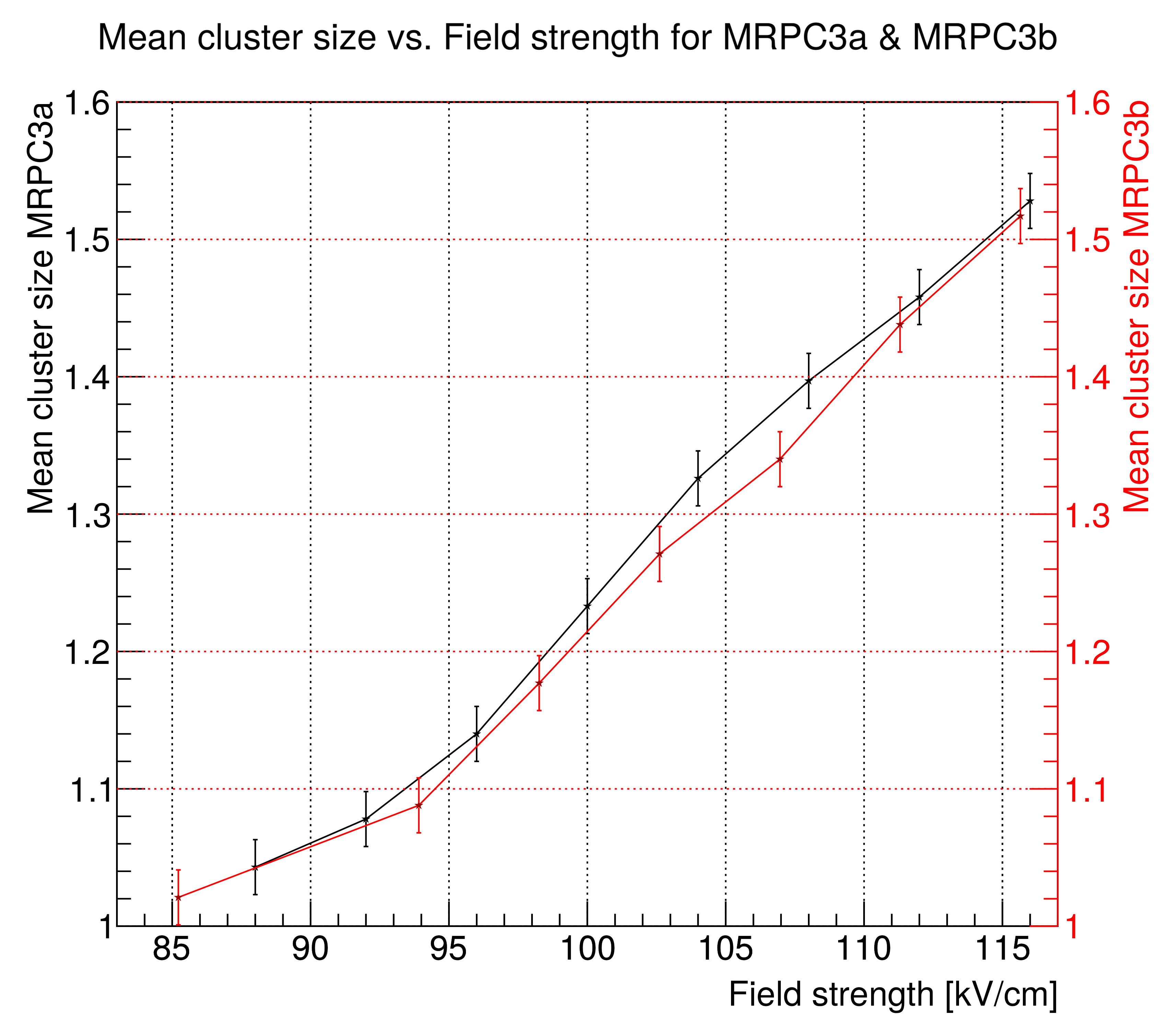}
	\qquad
	\includegraphics[width=.4\textwidth]{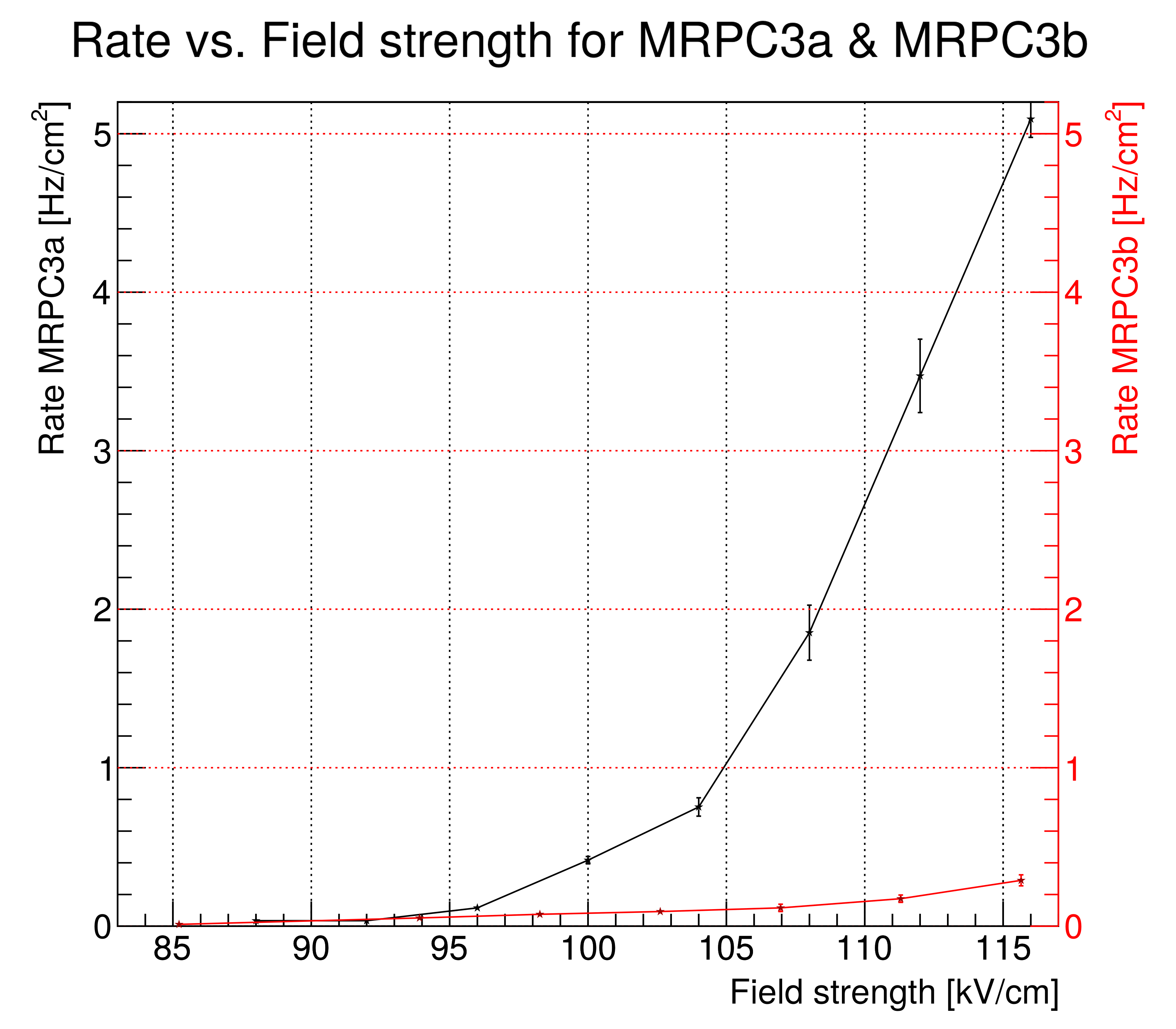}
	\caption{\label{fig:6} Left side: mean cluster size as a funtion of the el. field for the prototypes MRPC3a (low resistive glass) and MRPC3b (thin float glass). Right side: noise rate as a funtion of the el. field for the prototypes MRPC3a and MRPC3b.}
\end{figure}

Figure \ref{fig:5} shows the efficiency (left) and the counter time resolution (right) as function of the el. field for the MRPC3a and MRPC3b prototype. The efficiency of the MRPC3a is slightly lower than the one for MRPC3b which could be explained by the fact that MRPC3a has only 8 instead of 10 gaps. In addition MRPC3a has due to thicker glass plates a smaller weighting field. However both prototypes fulfill the CBM TOF requirements. The counter time resolution is constantly decreasing and reaches a value of about (50 $\pm$ 1)~ps for both prototypes which is expected since both MRPCs have similar gap sizes.
The cluster size (left) and the dark rate (right) as function of the applied el. field is depicted in fig. \ref{fig:6}. Since both MRPCs prototype have the same read out structure the cluster size is identical and rises linearly with the el. field strength. In the plateau the cluster size values are between 1.3 and 1.5. The noise/dark rate is, however, very different for both MRPCs. While for MRPC3b equipped with float glass electrodes the noise rate is very low ($\leq$ 0.3~Hz/cm$^2$) the noise rate for the MRPC3a equipped with low resistive glass electrodes rises stronger and roughly quadratically. At the operation voltage the value is more than a magnitude larger. The reason might be that the low resistive glass is polished and the surface is not as homogeneous as the one of float glass. However, a noise rate of 5~Hz/cm$^2$ is still tolerable since it is on the per mill level of the foreseen working rate.

\section{The FAIR phase 0 program of CBM TOF}
\label{sec:FAIR_phase0}

The FAIR phase 0 program is intended to install and operate detectors into existing experiments. CBM-TOF is involved in two projects:\\
1. The installation of 36 CBM TOF modules (about 7000 readout channels) at the east pol end cap at the STAR experiment at Brookhaven National Laboratory (BNL) extending the PID capability to the region of -1.5 $\leq$ $\eta$ $\leq$ -1.0. The so called eTOF wheel will be operational during the beam energy scan (BESII) campaign starting in February 2019.\\
2. The installation of 5 modules (25 MRPCs) in the miniCBM setup which is currently set up at SIS18 at GSI. The goal of this experiment is the systematic synchronization of the data flow and the data processing of the individual subsystems with the ultimate goal of $\Lambda$ reconstruction at sub-threshold energies. In addition it opens the possibility for detector test at rates that are anticipated for the running conditions at SIS100.\\

\begin{figure}[htbp]
    	\centering 
    	\includegraphics[width=.25\textwidth]{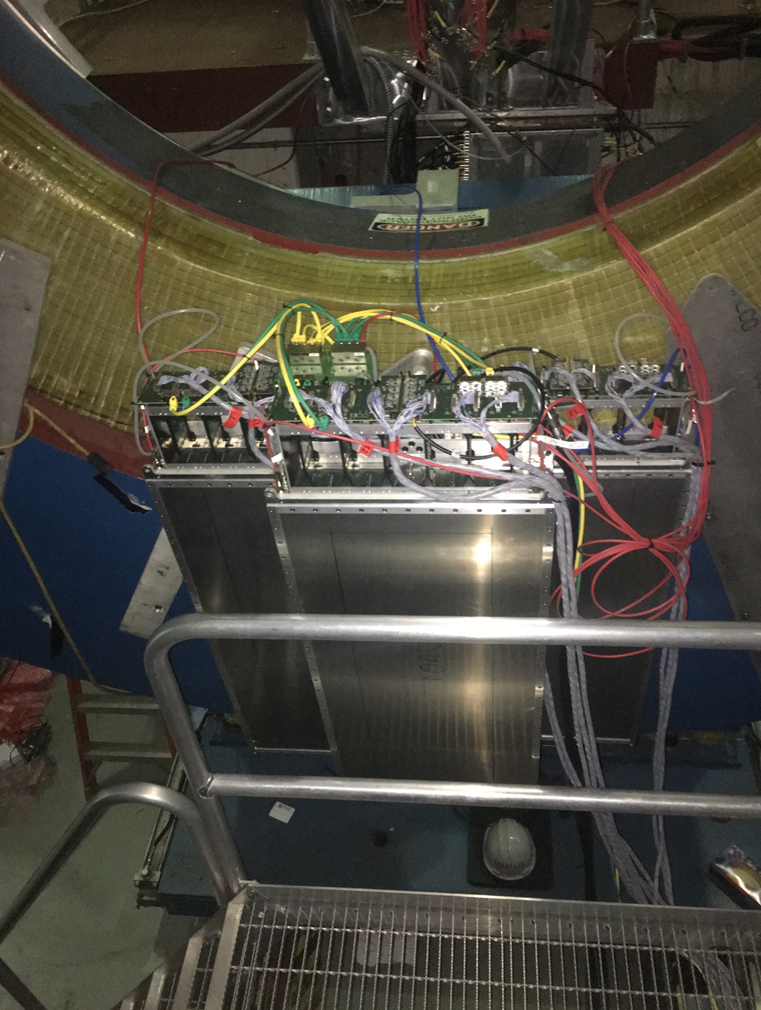}
    	\qquad
    	\includegraphics[width=.60\textwidth]{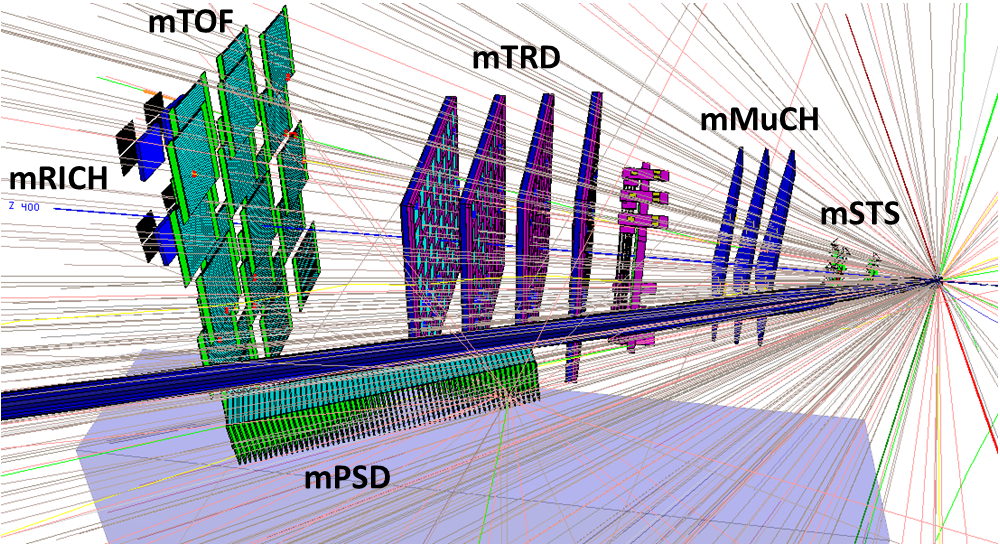}
    	\caption{\label{fig:7} Left side: Photograph of one sector installed at the STAR experiment at BNL. Right side: miniCBM setup with the several CBM subsystem.  }
    	\end{figure}

The left side of figure \ref{fig:7} depicts one sector (3 CBM TOF modules) mounted at the 6 o'clock position on the east side pole tip of the STAR solenoid. This sector was installed beginning of 2018 and tested during the STAR run18 beam time (February - June). Test results demonstrated the successful integration of CBMs free streaming DAQ systems into the triggered environment of the STAR experiment. The right hand side of figure \ref{fig:7} shows the mCBM setup that will be operational in Sep 2018.       

\section{Conclusions}
\label{sec:conclusions}

The CBM Time-of-Flight systems has to cope with unprecedented particle fluxes of up to 100 kHz/cm$^2$. Such conditions require new electrode materials for MRPCs which triggered the development of a new MRPC generation. In this paper we presented the possible solutions for CBM. Counter time resolution in the order of 50~ps were measured at low and moderate rates. With the opportunity to test in the high rate environment at SIS18 (mCBM) end of this year and the experience gained at BNL the design of the counter will be finalized and the mass production can start end of next year. The CBM TOF systems are targeted to be ready for operation end of 2023.    

\acknowledgments

Our thanks goes to all CBM TOF members for their great help to get this project running. We also thanks the German funding agencies: BMBF 05P15VHFC1



\end{document}